\documentclass[aps,prd,preprint,showpacs,amsmath,amssymb]{revtex4}
\usepackage{epsfig}

\newlength{\www}

\newcommand{\be}{\begin{equation}}
\newcommand{\ee}{\end{equation}}
\newcommand{\ba}{\begin{eqnarray}}
\newcommand{\ea}{\end{eqnarray}}

\newcommand{\bq}{\begin{equation}}
\newcommand{\eq}{\end{equation}}
\newcommand{\bqa}{\begin{eqnarray}}
\newcommand{\eqa}{\end{eqnarray}}
\newcommand{\ben}{\begin{enumerate}}
\newcommand{\een}{\end{enumerate}}
\newcommand{\bc}{\begin{center}}
\newcommand{\ec}{\end{center}}
\newcommand{\bqb}{\begin{eqnarray*}}
\newcommand{\eqb}{\end{eqnarray*}}

\begin{document}

\preprint{PTA/06-14}

\title{\vspace{1cm}
Single top production in the $t$-channel at LHC: a realistic test of electroweak models
}

\author{M. Beccaria$^{a,b}$, G. Macorini$^{c, d}$
F.M. Renard$^e$ and C. Verzegnassi$^{c, d}$ \\
\vspace{0.4cm}
}

\affiliation{\small
$^a$Dipartimento di Fisica, Universit\`a di
Lecce \\
Via Arnesano, 73100 Lecce, Italy.\\
\vspace{0.2cm}
$^b$INFN, Sezione di Lecce\\
\vspace{0.2cm}
$^c$
Dipartimento di Fisica Teorica, Universit\`a di Trieste, \\
Strada Costiera
 14, Miramare (Trieste) \\
\vspace{0.2cm}
$^d$ INFN, Sezione di Trieste\\
$^e$ Laboratoire de Physique Th\'{e}orique et Astroparticules,
UMR 5207\\
Universit\'{e} Montpellier II,
 F-34095 Montpellier Cedex 5.\hspace{2.2cm}\\
\vspace{0.2cm}
}

\begin{abstract}
We compute the complete electroweak one-loop effect on the process of           
$t$-channel single top production at LHC in the Standard Model and in the MSSM
within the mSUGRA symmetry breaking scheme. We find that the one-loop electroweak SM
effect is large, and decreases the cross section of an amount that is of the
same size as that of the NLO QCD one. The genuine SUSY effect in the mSUGRA
scheme, for a general choice of benchmark points, is rather small. 
It might become large and visible in more general scenarios around thresholds
involving light stop and neutralino mass values. 
\end{abstract}

\pacs{12.15.-y, 12.15.Lk, 13.75.Cs, 14.80.Ly}

\maketitle

\section{Introduction}
\label{sec:intro}

The relevance of the process of single top production at LHC has been 
already stressed by
several authors~\cite{Tait:1999cf,Stelzer:1997ns,Mahlon:1999gz}. 
In the Standard Model (SM) framework, it appears as  a unique way of  measuring the $tbW$ 
coupling that appears already at the Born level of the scattering 
amplitude. Deviations  from the unitarity value would be indications of 
New Physics, essentially of different kind in the three basic processes 
that are involved in the single top production, currently named $t$-channel or $Wg$ fusion, 
associated production and $s$ channel reactions. 
Roughly, any 
deviation from the expected CKM prediction $V_{tb}\simeq 1$ would 
decrease the SM cross section by the amount $|V_{tb}|^2$ for all the 
three processes.
Clearly, the presence of three simultaneous decreases would therefore be 
a strong indication for
the presence of  some particular form of physics beyond the SM, leading 
to a violation of the CKM prediction. Alternatively, though, one might 
discover deviations from the SM predictions
affecting the three processes in a different way, for instance if an extra 
$W'$ boson existed, which would only affect the $s$-channel process but not 
the two other ones. Also, supersymmetric virtual exchanges at one loop 
might produce sizable and in principle not identical effects in 
the processes. Given the ambitious final goal of the LHC top quark working 
group of measuring the various cross sections with an overall uncertainty of $5\%$~\cite{Yellow}, 
any relevant theoretical proposal should 
start from the observation of effects whose numerical size represents a 
deviation from the SM value of, at least, that value.

For what concerns the SM predictions, the picture nowadays is, in our 
opinion, the following one. Numerically, the dominant process at LHC, {\em i.e.} 
that  with the largest cross section, is the $t$-channel one 
for which a value of approximately 245 pb (total cross section) 
is expected. The second process is that of associated production 
with an expected value of approximately 60 pb. Third, the s 
channel process with a value of  about 10 pb. 
Although this might be a too drastic attitude, we believe that a 
theoretical estimate of the SM prediction beyond the simple  Born 
approximation  would be requested for the two previous processes, but 
not for the third one,  at least in a preliminary phase of the LHC 
activities. In this spirit, from now on, we shall concentrate our attention 
on these two dominant reactions and summarize the status of the SM 
calculations.

A priori, one expects that the dominant NLO effect is that due to QCD 
corrections. The latter ones have been evaluated in~\cite{QCDNLO}, including 
NLO top quark decay~\cite{QCDNLODecay}, and matching the NLO matrix elements 
to the parton shower framework~\cite{Frixione:2005vw}.
The typical (cut dependent) overall relative effect is of approximately $+10\%$
for the associated $tW$ production process and of an absolute magnitude of a few percents
in the $t$-channel process. In a sense, these sizes appear to us "optimal" in 
the sense that they are "sufficiently mild" to prevent the need of 
higher order QCD effects, and at the same time "sufficiently strong" to 
be kept into account at a hopeful 5\% error level.

For what concerns the electroweak effects, the first (and unique, to our 
knowledge) complete
one-loop calculation for the associated production has been performed by 
our group~\cite{Beccaria:2006dt} in the specific case of the MSSM.  Briefly, 
the main (essentially negative) result is that the genuine SUSY effect 
is generally small, of the few percent at most, for various typical 
choices of the benchmark points. The SM effect on the total cross 
section is also small at one loop, although it might reach a 10\%
relative value at high energies (of the one TeV size), where though it is   
not evident that the process may be distinguished from the dominant $t\overline t$
background.

A feature that appears to us worth being mentioned appears in the 
calculation of the electroweak SM one-loop effect for $tW$ production. In 
a previous preliminary paper~\cite{Beccaria:2004xk} we gave an 
approximate estimate supposedly valid for energy values sufficiently 
larger that those of the masses of the (real and virtual) particles 
appearing in the various Feynman diagrams. This estimate was based on a 
logarithmic expansion of Sudakov kind, computed to logarithmic NLO
i.e. only retaining the squared and the linear 
logarithms of the expansion. Comparing the complete SM  one-loop 
calculation with this approximation we found that for c.m. energies 
beyond, roughly, 500 GeV the result was essentially identical with 
the approximate Sudakov expansion. This would be possibly relevant e.g. to 
prepare a simplified effective parametrization if the process were still 
observable at those energies, which is not clear at the moment.

Numerically speaking, one sees that the SM effect would be, beyond 500 GeV, always 
negative and of a size increasing from  roughly 5\% to 10\%
at the assumed limit of our analysis of 1 TeV. Given the aforementioned
features of the process, this does not appear to be a particularly relevant
result for the purposes of a SM precision test, at least for the moment.

The motivations of this paper are a result of the same analysis performed in~\cite{Beccaria:2004xk}. 
From Fig.~(4) of that reference, one sees that, in the c.m. energy 
region beyond 500 GeV, that now appears experimentally valid, the
Sudakov expansion of the distribution exhibits a SM effect that is largely
beyond the 10\%, reaching a final value of approximately
25\% at the assumed limit of analysis of 1 TeV. If this feature 
persisted, at least partially, in a complete calculation, it would provide the 
opportunity of a realistic precision test at LHC of the electroweak component of the 
SM, and possibly of other models of electroweak physics whose one-loop effect
were of the same respectable size.

With these premises, we present in this paper the first complete one-loop
analysis of the electroweak effect on the $t$-channel process in the MSSM, 
in the mSUGRA scenario of symmetry breaking. From our previous experience with
the $tW$ production process, we do not expect to discover in this scheme, for realistic values
of the sparticle masses, exciting genuine
SUSY effects, like those that we found at logarithmic NLO in the specially light 
SUSY situation considered in~\cite{Beccaria:2004xk}. 
Independently of that, we shall pay a special attention to the
pure SM effect, since values for the latter consistent with the Sudakov
predictions should certainly be carefully retained in any search of new physics
that aims to be compared with experiments having an overall error of the 5\% size.

Technically speaking, the paper will be organized as follows:
Section 2 will be devoted to a short illustration of the
cancellation of ultraviolet and infrared divergences, keeping in mind the fact 
that several details have already been thoroughly discussed in~\cite{Beccaria:2006dt} and will 
be therefore treated as concisely as possible. In the first part of Section 3 we shall 
discuss the pure SM effect and its strong connections with the Sudakov
approximation; In the second part, the genuine SUSY effect will be shown for several 
choices of benchmark points. Given its generally small size, a tentative and 
qualitative illustration of a possibly large effect for very special choices 
of the SUSY parameters, in particular corresponding to one light stop mass,
(in a general MSSM point), will be shown. In section 4 some possible conclusions of our
work will be drawn.

\section{Electroweak MSSM $ub\to td$ production at one loop}

We shall now describe the one-loop description of the process $ub\to td$. 
We shall renormalize the process according to the {\em on-shell} 
renormalization scheme. Our notation will be consistent with~\cite{Hollik:1988ii}.
We shall systematically neglect radiative corrections proportional to powers of the light quarks masses.
This approximation simplifies the calculation and, for instance,  allows to drop all diagrams with 
a propagation in the $t$-channel of a virtual particle different than $W^-$.

At Born level, the process $ub\to td$ is described by a single diagram describing $W$ propagation in the $t$-channel.
We define the Mandelstam variables
\be
s=(p_b+p_u)^2=(p_t+p_d)^2, \ t=(p_b-p_t)^2=(p_u-p_d)^2, 
\ee
and also introduce $q'=p_b-p_t=p_d-p_u$, so $t=q'^2$. The Born amplitude is 
\bqa
A^{Born}=\frac{e^2}{2s^2_W(t-M^2_W)}
[\bar u(t)\gamma^{\mu} P_Lu(b)][\bar u(d)\gamma_{\mu} P_Lu(u)]
\eqa
This Born term receives radiative corrections that can be split in several classes. In details, they consists of 
(i) counterterms and internal/external self-energies
contributions, (ii) vertex corrections to the light or heavy quark charged current, (iii) box contributions, i.e. genuine one-particle
irreducible four legs diagrams, (iv) real soft photon radiation.
We shall now discuss separately each  class and 
its specific features and, after that, we shall make some comments about the high-energy
expansion of the process which is known analytically at NLO in the large logarithms and
offers a non trivial check of the calculation.


\subsection{Self-energies and counterterms}

The counterterms and self-energy contributions can all be expresses in terms of the 
external quarks and gauge bosons self energies~\cite{Hollik:1998md}.
The amplitude correction due to self-energies and counterterms reads
\bqa
A&=&A^{Born}~[~1+2(\delta Z^W_1-\delta Z^W_2)\nonumber\\
&&+{\frac 1 2}
(\delta Z^b_L+\delta Z^u_L+\delta Z^t_L+\delta Z^d_L
+\delta\Psi_t+\delta\Psi_u)
-{\hat{\Sigma}^{WW}(t)\over t-M^2_W}~]
\eqa
where the fermionic counterterm contributions are  given in terms of
"down quark" quantities\cite{Hollik:1988ii} $f=d,b$, exploiting $SU(2)_L$ symmetry. As a 
consequence the residue of the up-type quark propagator at the mass pole must be corrected by
adding a finite wave-function renormalization relative to the up-type light and heavy quarks.

\bq
\delta Z^t_L=\delta Z^b_L~~~~~~\delta Z^u_L=\delta Z^d_L
\eq

with for $f=d,b$

\bq
\delta Z^f_L=
-\Sigma^f_L(m^2_f)-m^2_f[\Sigma^{'f}_L(m^2_f)+\Sigma^{'f}_R(m^2_f)
+2\Sigma^{'f}_S(m^2_f)]
\eq

\bq
\delta\Psi_u=-\{\Sigma^u_L(m^2_u)+\delta Z^d_L+
m^2_u[\Sigma^{'u}_L(m^2_u)+\Sigma^{'u}_R(m^2_u)
+2\Sigma^{'u}_S(m^2_u)]\}
\eq
\bq
\delta\Psi_t=-\{\Sigma^t_L(m^2_t)+\delta Z^b_L+
m^2_t[\Sigma^{'t}_L(m^2_t)+\Sigma^{'t}_R(m^2_t)
+2\Sigma^{'t}_S(m^2_t)]\}
\eq
and with (see also Hollik lectures)

\bq
\delta Z^W_1-\delta Z^W_2=~{\Sigma^{\gamma Z}(0)\over s_Wc_W M^2_Z}
\eq

\bqa
{\hat{\Sigma}^{WW}(t)\over t-M^2_W}&=& 
{\Sigma^{WW}(t)-Re\Sigma^{WW}(M^2_W)\over t-M^2_W}\nonumber\\
&&-\Pi^{\gamma}(0)-2{c_W\Sigma^{\gamma Z}(0)\over s_W M^2_Z}
+{c^2_W\over s^2_W}({Re\Sigma^{ZZ}(M^2_Z)\over M^2_Z}-
{Re\Sigma^{WW}(M^2_W)\over M^2_W})
\eqa

Concerning UV divergences, we remark that the renormalized function $\hat{\Sigma}^{WW}$
is convergent but the unrenormalized functions $\Sigma$ are generally divergent. 
The resulting divergences in $A$ due to the combination
$2(\delta Z^W_1-\delta Z^W_2)+\delta Z^b_L+\delta Z^d_L$
will be canceled by the divergences appearing in the vertex corrections. This works
separately for the purely SM and genuine SUSY subsets of diagrams.

\subsection{Vertex corrections}

The vertex corrections are the one-loop diagrams correcting the 
charged currents associated to the light or heavy quarks. We simply list the relevant 
classes of diagrams, i.e. denote subclasses of diagrams by the internal (possibly generic) 
virtual particles.

For the light quark charged current, we computed the following six classes of diagrams
\be
(uV^0d),\
(WdV^0),\
(V^0uW),\
(\chi^-_i\tilde d_{L,R} \chi^0_j),\
(\chi^0_j\tilde u_{L,R} \chi^+_i),\
(\tilde u_{L} \chi^0_j\tilde d_{L}),
\ee
where $V^0 = \gamma, Z^0$.

For the heavy quark charged current, we have instead 13 classes
\be
(tV^0b),\ 
(tSb),\ 
(WbV^0)  ,\ 
(V^0tW)  ,\ 
(V^0tS^-),\ 
(WbS^0)  ,\ 
(S^+bV^0),
\ee
\be
(S^0tW)  ,\ 
(S^0tS^-),\ 
(S^+bS^0),\ 
(\chi^+_i \tilde b_{L,R}\chi^0_j ),\ 
(\chi^0_j \tilde t_{L,R}\chi^-_i ),\ 
(\tilde t_{L}\chi^0_i \tilde b_{L}).
\ee
where $V^0 = \gamma, Z^0$, and $S^{0, \pm}$ denote a neutral or charged scalar particle.

\subsection{Box corrections}

\begin{figure}[tb]
\centering
\epsfig{file=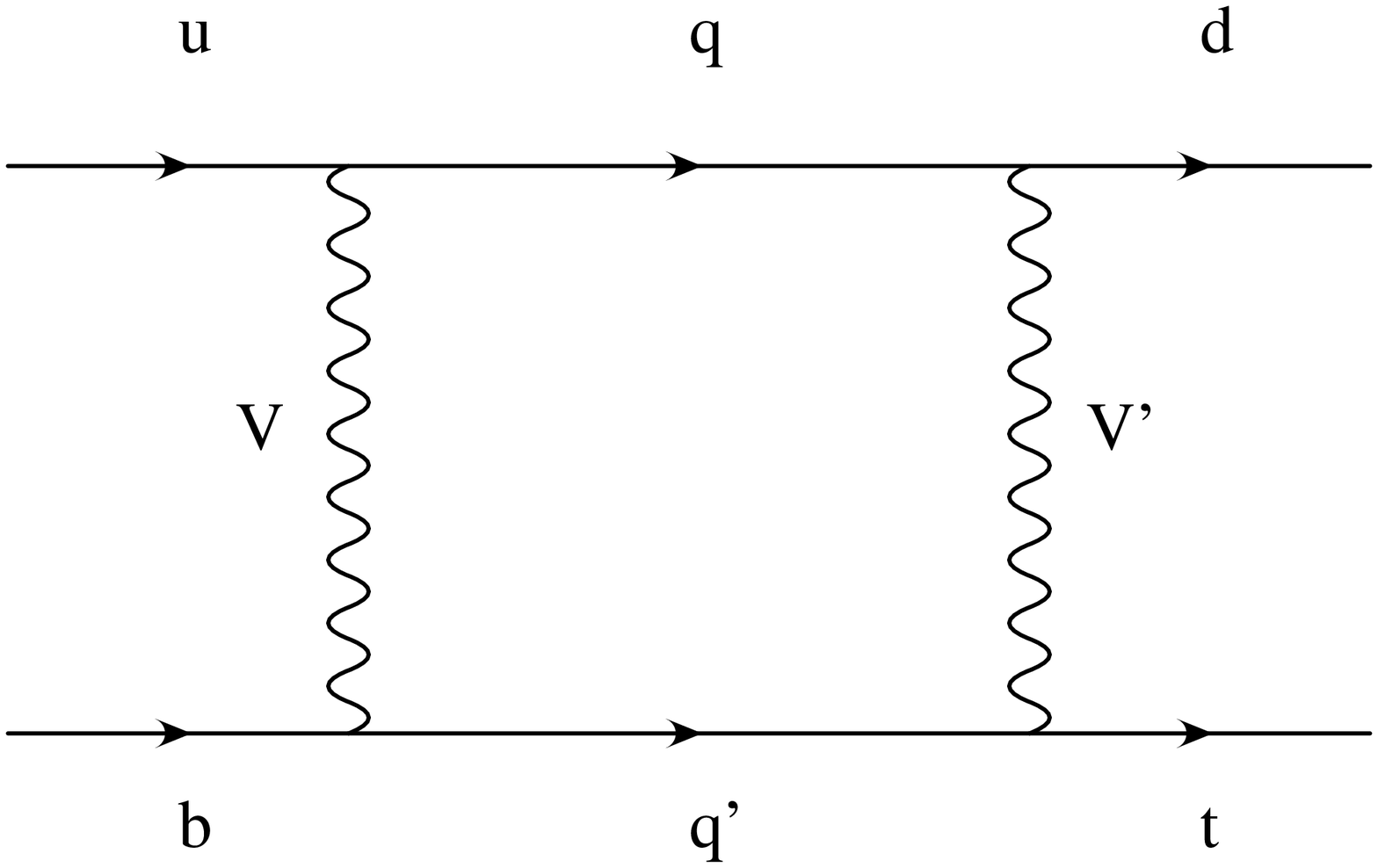, width=6cm, angle=0}
\ \ \ \ 
\epsfig{file=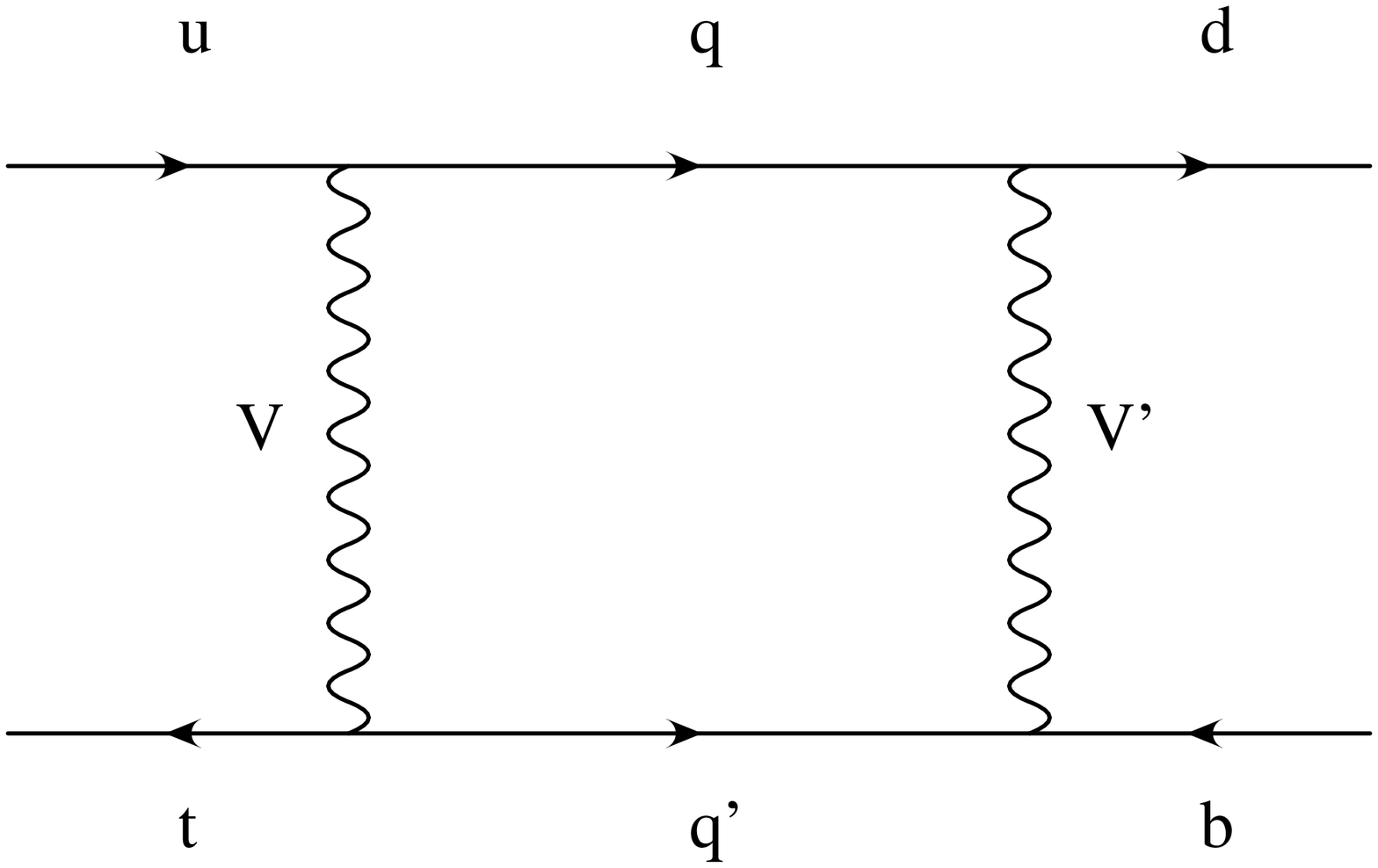, width=6cm, angle=0}
\vspace{1.5cm}
\caption{Standard Model direct and twisted box diagrams. The virtual $q$ and $q'$
are quarks. The gauge bosons $(V, V')$ can be $(\gamma,W)$, $(Z,W)$, $(W,\gamma)$ or $(W,Z)$.}
\label{fig:boxesSM}
\end{figure}

\begin{figure}[tb]
\centering
\epsfig{file=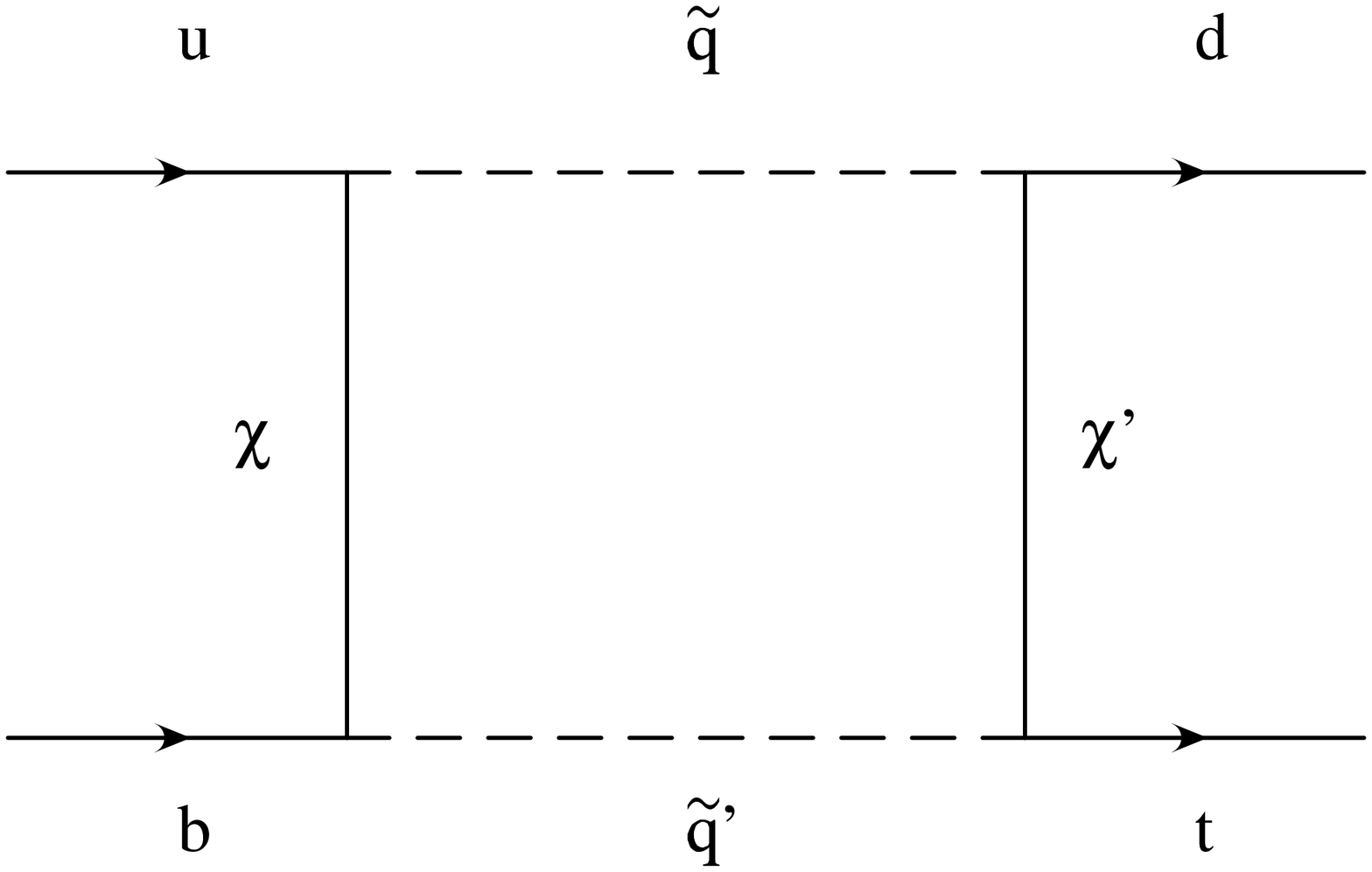, width=6cm, angle=0}
\ \ \ \ 
\epsfig{file=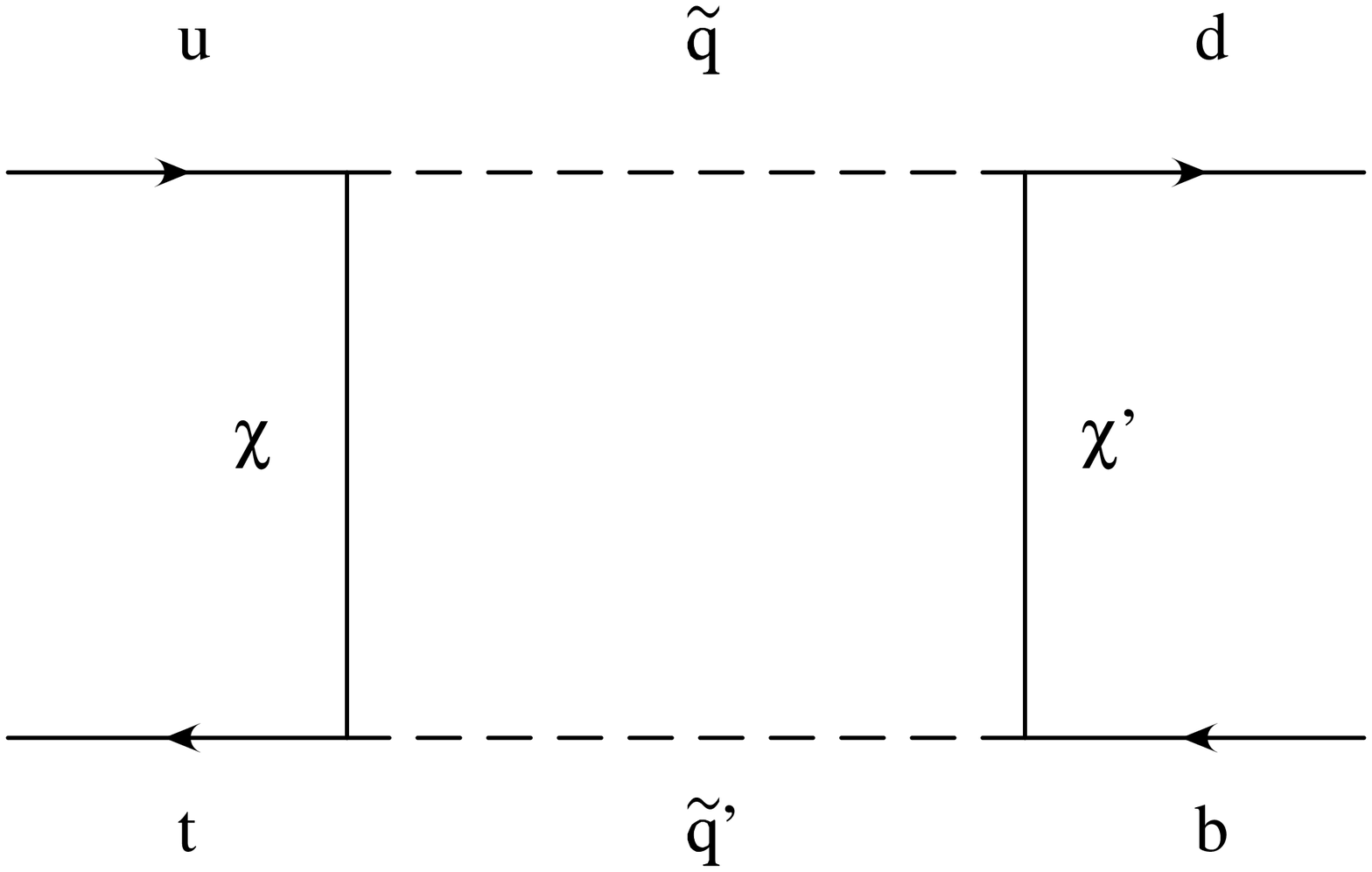, width=6cm, angle=0}
\vspace{1.5cm}
\caption{SUSY direct and twisted box diagrams. The virtual $\widetilde{q}$ and $\widetilde{q}'$
are squarks. The fermion lines $(\chi, \chi')$ can be charginos or neutralinos,
$(\chi^0, \chi^+)$ or $(\chi^+, \chi^0)$.
}
\label{fig:boxesSUSY}
\end{figure}

We considered four classes of box diagrams. They are shown in Fig.~(\ref{fig:boxesSM}, \ref{fig:boxesSUSY}).
As a general remark, we remind that box diagrams are not UV divergent in this process. 
Of course, those with the exchange of a virtual photon 
produce hard IR divergences to be canceled by the real soft radiation, as usual.

\subsection{Real photon radiation and IR finiteness}

QED radiation effects are usually split into a soft part containing the potential IR singular terms, and a 
hard part including the emission of photons with energy not small compared to the process energy scale.
In this brief section, we only discuss the soft emission and the detailed cancellation of IR divergences
that occurs when it is combined with virtual photon exchanges.

We denote by ${\cal A}^{\rm Born}$ and ${\cal A}^{\rm 1\ loop}$ any invariant helicity scattering amplitude
evaluated at Born or one loop level. The IR regulating fictitious photon mass will be denoted by $\lambda$.
The IR cancellation between (soft) real radiation and virtual photon exchange holds in every helicity channel separately and we 
have checked it numerically. It reads
\be
\left({\cal A}^{\rm Born}\right)^2 \left(1 + \frac{\alpha}{2\pi}\delta_s\right) + 2 {\cal A}^{\rm Born}\ {\cal A}^{\rm 1\ loop} = \mbox{IR finite}.
\ee
Here, $\delta_S$ is the correction factor taking into account the emission of soft real 
photons with energy from $\lambda$ up to $E_\gamma^{\rm max} \ll \sqrt{s}$~\cite{'tHooft:1978xw}.
In fact, the singular part of $\delta_S$ is quite simple
\be
\delta_S = \log\frac{\lambda}{E_\gamma^{\rm max}}\ \sum_{i,j} \delta_S^{i,j}   + \mbox{regular terms as}\ \lambda\to 0
\ee
where $i$ and $j$ runs over all pairs of external particles. There are two types of contributions $\delta_S^{i,j}$:
The diagonal ones with $i=j$ and the off diagonal ones with $i\neq j$~\cite{'tHooft:1978xw}.
The diagonal terms with $i=j$ match the IR divergence in the counterterms
associated to the $i$-th external line~\cite{Weinberg:1995mt}.
The  off-diagonal radiation terms $i\neq j$ match the IR divergence in the diagrams
where the $i$-th and $j$-th external lines are connected by a virtual photon. These can be of vertex or box type.
For our preliminary analysis, we have fixed the (reasonable)
value $E_\gamma^{\rm max} = 0.1$ GeV. A more complete analysis, that takes into account the 
effects of hard photon radiation, will be presented in a separate forthcoming
paper~\cite{forthcoming}.

\subsection{High-energy behavior}

The high energy behavior of the $ub\to td$ process is known analytically at NLO order in the 
Sudakov expansion~\cite{Beccaria:2004xk}. It has been derived according to general rules for the Sudakov 
expansion of SM or MSSM processes. In the Appendix, we shall recall these results and show in some details 
how they can be recovered from the explicit one-loop diagrammatic expansion. 
We shall also give in the final Section  a detailed discussion of applicability of this expansion  and of the 
matching between the high and low energy regimes.

\section{Physical Predictions}

We shall concentrate our analysis on the investigation of the electroweak one-loop
MSSM effect on the unpolarized cross section, for which a preliminary 
discussion of the expected experimental error already exists~\cite{Yellow}.
In principle, the final top polarization could also be measured, but a similar
experimental analysis has not yet been completed, to our knowledge.
The starting quantity will be therefore the inclusive differential cross section
of the process, defined as usual as:
\ba
\label{eq:basic}
{d\sigma(PP\to t d+X)\over ds}&=&
{1\over S}~\int^{\cos\theta_{max}}_{\cos\theta_{min}}
d\cos\theta~[L_{ub}(\tau, \cos\theta)
{d\sigma_{ub\to  t d}\over d\cos\theta}(s)~]
\ea
\noindent
where $\tau={s\over S}$, and $L_{ub}$ is the parton process luminosity.
\be
L_{ub}(\tau, \cos\theta)=
\int^{\bar y_{max}}_{\bar y_{min}}d\bar y~ 
~\left[~ b(x) u({\tau\over x})+u(x)b({\tau\over x})~\right]
\label{Lij}
\ee
\noindent
where S is the total pp c.m. energy, and 
$i(x)$ the distributions of the parton $i$ inside the proton
with a momentum fraction,
$x={\sqrt{s\over S}}~e^{\bar y}$, related to the rapidity
$\bar y$ of the $td$ system~\cite{QCDcoll}.
The parton distribution functions are the latest LO MRST (Martin, Roberts, Stirling, Thorne) 
set available on~\cite{lumi}.
The limits of integrations for $\bar y$ depends on the cuts. We have chosen a 
maximal rapidity $Y=2$ and a minimum $p_T$ which we shall specify later.

Note that we are at this stage considering as kinematical observable the 
initial partons c.m. energy $\sqrt{s}$, and not the realistic final 
state invariant mass $M_{td}$. 

To relate these two 
quantities requires a straightforward analysis that will be performed in the
announced more complete forthcoming paper~\cite{forthcoming}. We expect from
our previous investigation performed for the process of $t\overline{t}$ annihilation~\cite{Beccaria:2004sx}
that the difference between $M_{td}$ and $\sqrt{s}$
is relatively small, not beyond the 
relative 5\% level, particularly in the energy region that we shall
consider. In the following part of the paper we shall therefore provide 
plots of quantities at variable $\sqrt{s}$. In the first part of the Section,
the SM result will be discussed.

\subsection{SM results}

\begin{figure}[tb]
\centering
\epsfig{file=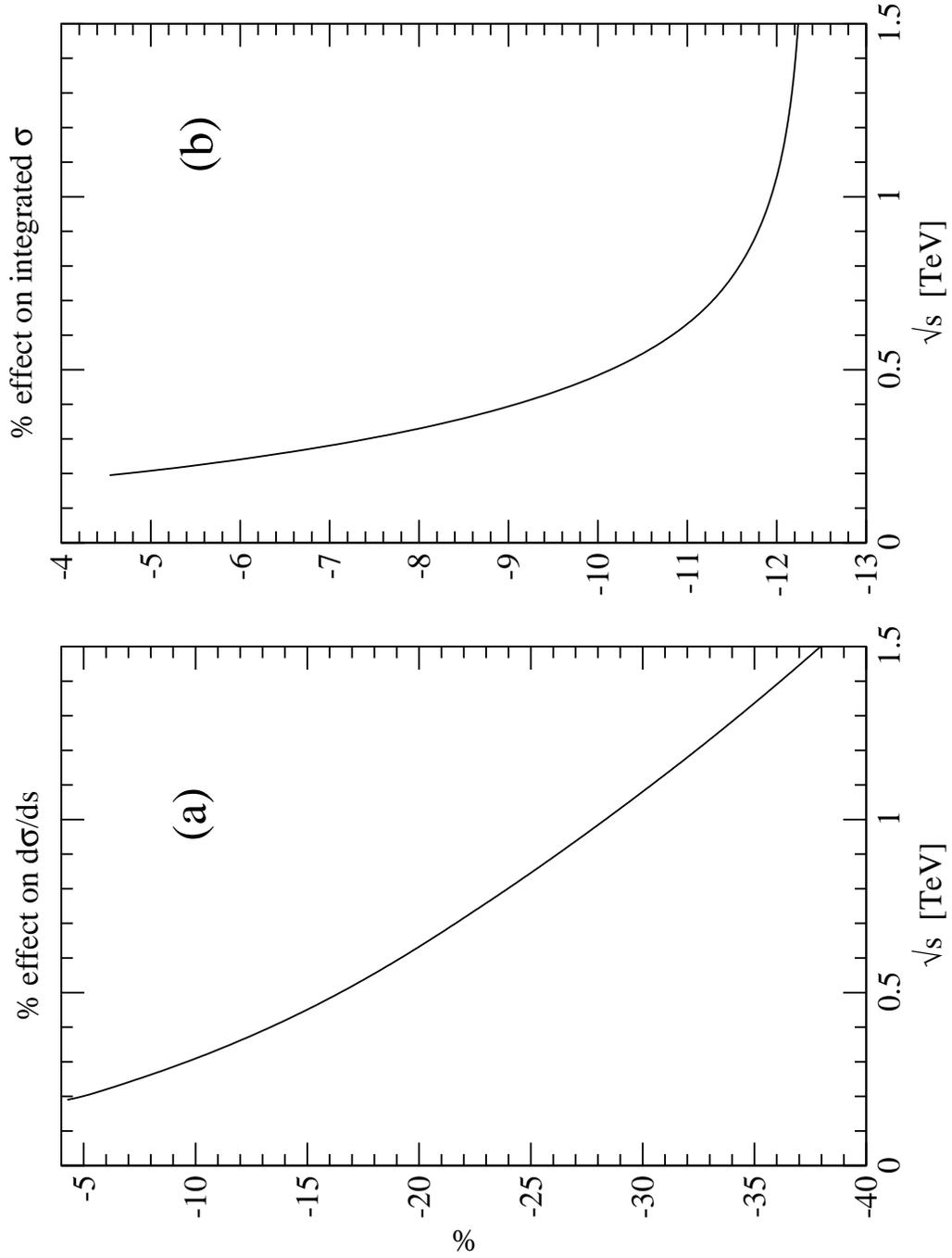, width=18cm, angle=90}
\vspace{1.5cm}
\caption{Full one loop calculation in the Standard Model. Panel (a) shows the percentual effect on the distribution $d\sigma/ds$.
Panel (b) shows the effect on the integrated cross section from threshold up to $\sqrt{s}$.}
\label{fig:SMEffects}
\end{figure}

We begin the presentation of our results with the standard model case. To compute
all the physical quantities, we have written a C++ numerical code available upon request. It passes all the checks that have been discussed, 
{\em i.e.} cancellation of UV and IR divergences and correct high-energy behavior. 
In Fig.~(\ref{fig:SMEffects}, a) we show the percentage one-loop effect for the
differential distribution $d\sigma/ds$ having used the values $p_{T, \rm min} = 10$ GeV.
The effect is always negative and increases in magnitude with energy up to quite large values. 
Of course the total integrated cross section is dominated by moderate values of $\sqrt{s}$ not much 
larger than the production threshold. Hence, to appreciate the actual relevance of the effect, we show in 
Fig.~(\ref{fig:SMEffects}, b) the percentage one-loop effect on the integrated cross section 
from threshold up to a certain $\sqrt{s}$. The curve saturates around 700-800 GeV where it reaches a 
plateau effect of about -12\%.

\begin{figure}[tb]
\centering
\epsfig{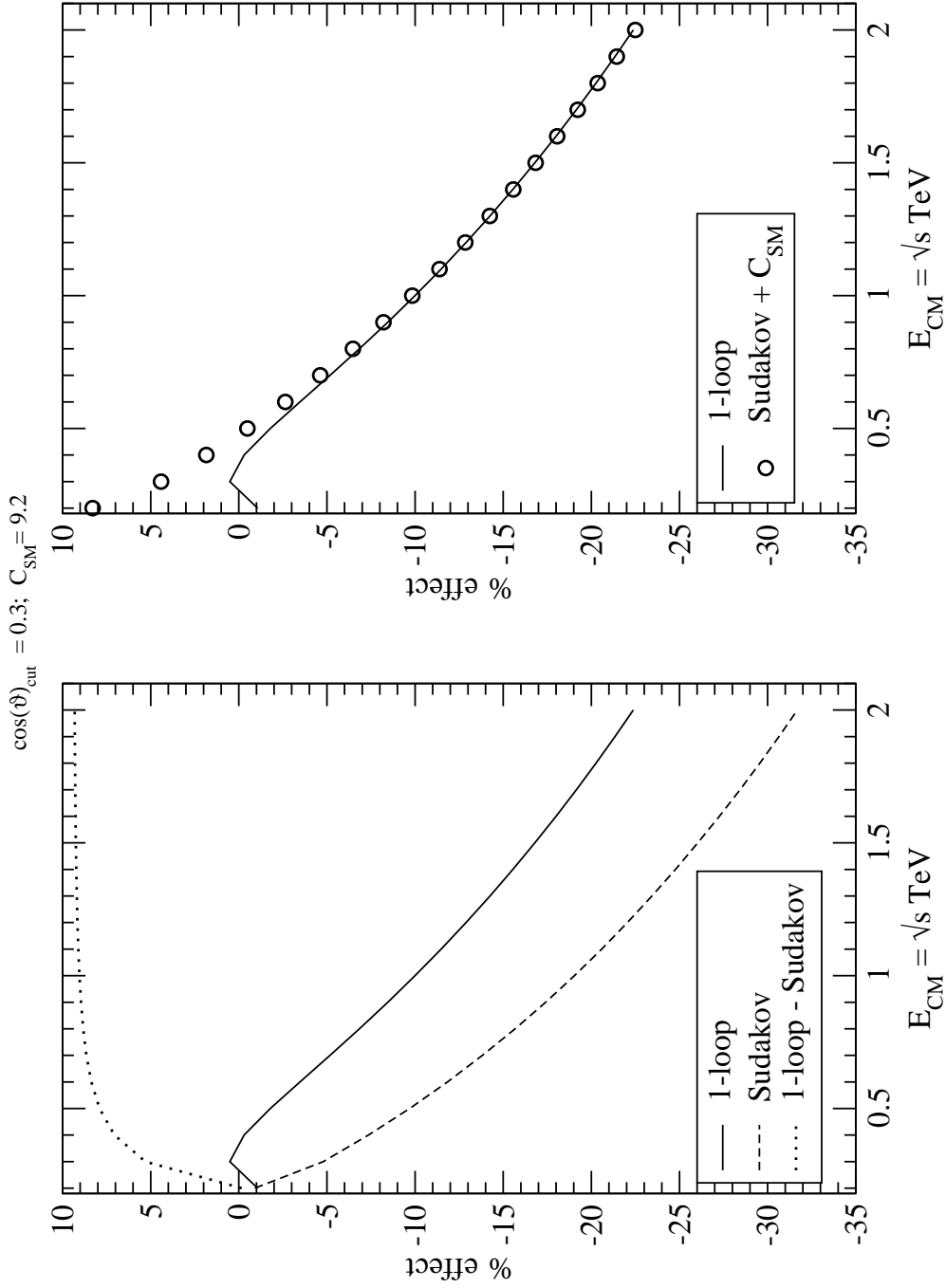}
\vspace{1.5cm}
\caption{Comparison between the full one-loop calculation in the Standard Model and the NLO Sudakov approximation.
As explained in the text, a strong angular cut is imposed as well as the fictitious definition $M_\gamma = M_Z$. Real QED radiation
is consistently switched off.}
\label{fig:sudakovSM}
\end{figure}

The full one-loop effect can be compared with the Sudakov approximation. With this aim, we fix a suitable 
kinematical configuration. In particular, we impose a strong angular cut to avoid the region of small $t$
which is physically the most important, but where the Sudakov approximation fails since it requires
$s, t, u$ to be much larger than the process typical mass scales. Also, for the purpose of comparison, we 
switch off the QED real corrections and regulate the IR divergent one loop diagrams with the fictitious mass
$M_\gamma \to M_Z$. As explained in the Appendix, this is needed in order to exploit the $SU(2)\times U(1)$
inspired simple expressions for the Sudakov corrections. The result of the comparison is shown in 
Fig.~(\ref{fig:sudakovSM}). Here, one can see that just above 500 GeV, the NLO Sudakov approximation provide
a quite good representation of the energy slope of the distributions.
In practice, a fitted constant representing the NNLO term in the expansion is enough to reproduce quite
accurately the full one-loop result with a value of the constant that reduces the effect by an amount that
approaches, at 1 TeV, the 50\% of the logarithmic approximation. A similar discussion and conclusion can be found in the recent 
paper~\cite{Moretti:2006nf} computing SM electroweak corrections to the process $gg\to t\overline{t}$ at LHC
and also in old analyses of our group\cite{Beccaria:1999xd}.

\subsection{MSSM results}

All the calculations of the one-loop effect in the MSSM case can been  performed
running our numerical code for various choices of the MSSM parameters. In this
preliminary analysis we have chosen the mSUGRA breaking scheme and retained 
its set of SUSY parameters, conventionally denoted $m_0$, $m_{1/2}$, $A_0$, 
$\tan\beta$, $\mbox{sign}\,\mu$. 

\begin{figure}[tb]
\centering
\epsfig{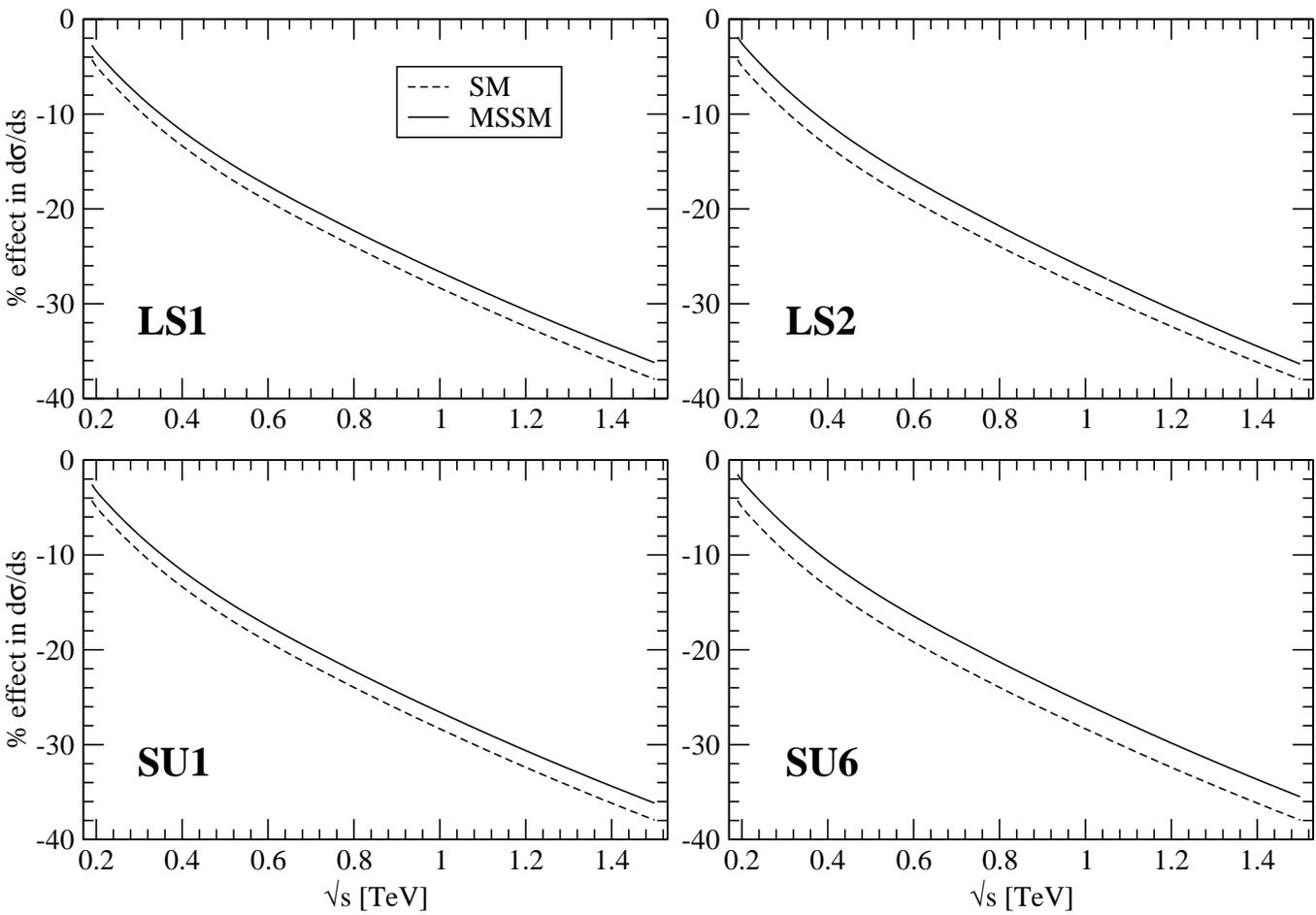}
\vspace{1.5cm}
\caption{Percentual effects in the differential cross section $d\sigma/ds$ in the four
considered MSSM scenarios. The SM curve is also shown in each case.}
\label{fig:mssmeffects}
\end{figure}

We have
examined several benchmark points already existing in the literature. As a 
general feature, we have found a relatively small genuine SUSY effect, typically
of the few percent size. In Fig.~(\ref{fig:mssmeffects}) we have shown the comparison between the SM
one-loop effect and the MSSM  one, having chosen four benchmark points that
correspond to different choices of the parameters, that are normally defined as
SU1, SU6, LS1, LS2. 
Two of them are the 
ATLAS DC2 SU1 and SU6 points~\cite{DC2}; the remaining two are two points 
whose spectrum has been evaluated by the code SUSPECT~\cite{Djouadi:2002ze} and that 
we have called LS1, LS2 where LS stands for Light SUSY.

In Table I we have listed the physical masses of
sparticles that correspond to the four choices. As one can see, the genuine SUSY
effect varies between, approximately, two and three percent, depending on the
chosen point. At the aimed LHC accuracy level of five percent for this process,
the SUSY effect in the mSUGRA scenario appears in general definitely too small for 
being detected, independently on the chosen values of the parameters. This
negative conclusion is the same that was derived by our previous analysis
of~\cite{Beccaria:2006dt}, and it deserves, we believe, a number of comments.

The first question concerns the big difference between the complete one-loop
calculation of this paper and the approximate Sudakov expansion given in~\cite{Beccaria:2004xk}. 
As a matter of fact, that analysis was performed 
assuming a {\em specially light} SUSY scenario, where all the sparticle masses
were assumed to be lighter than a few hundred GeV. The effective SUSY
mass $M_{\rm SUSY}$ that appeared in the logarithmic expansion in terms of $\log\frac{s}{M_{\rm SUSY}^2}$
was then
assumed to be of the same size, i.e. a few hundred GeV at most. In the mSUGRA
scheme, for all the benchmark points that we found, this scenario does  not
appear, as one can see from Table I. As a consequence, there are no longer
large (linear) SUSY logarithms $\log(s/M_{\rm SUSY}^2)$ of Yukawa kind, enhanced by large 
$\tan\beta$ values. Alternatively, one may think that these terms can still be
retained, with a fictitious light SUSY mass $M_{\rm light}$, but at the price of adding a 
potentially large next-to-next-to leading (i.e. energy independent) term $\log(M_{\rm heavy}/M_{\rm light})$
where $M_{\rm heavy}$ is the real effective SUSY scale. This term has
clearly opposite sign with respect to the fictitious one, and consequently
it manages to destroy it.

To investigate whether this simplified explanation is correct, we show in the
next Fig.~(\ref{fig:sudakovMSSM})
the comparison of the complete one-loop effect, in 
two of the four chosen benchmark points, with the
"fictitious"  Sudakov purely logarithmic approximation, done using a light
$M_{\rm SUSY}$ effective mass, of the one hundred GeV size.  
The comparison is done along the same lines discussed in the previous Section for the 
SM case.
As one sees, as soon as the
energy becomes larger than, approximately, five hundred GeV, the difference
between the two
calculations becomes, indeed, a constant (energy independent) term. This term
decreases the size of the logarithmic Sudakov approximation, of a relatively
 large amount that 
varies numerically in the four cases. One sees also that the size of the constant 
term is definitely larger in those cases where the sparticle masses are larger,
in agreement with the qualitative argument that we have given in the previous
discussion.

An almost unavoidable conclusion is that a potentially large genuine SUSY effect
necessitates a scenario where at least some of the virtual particles that can
be exchanged in the Feynman diagrams are, indeed, light, in particular with respect
 to the realistic energies of the process. From an experimental point of view,
an upper limit of energy could be placed in our opinion at about 1 TeV,
 and a realistic range to be examined
might be 500-1000 GeV. We shall assume for the moment
 realistic experimental conditions in this range. We do  not have yet at disposal
an accurate experimental analysis for this process, analogous to that that was
performed by members of the top Atlas group for the process of $t\overline{t}$
production~\cite{Beccaria:2004sx}. This analysis is actually in progress, and the results should
appear soon~\cite{forth2}.

A first possibility appears to be that of abandoning the mSUGRA symmetry
breaking scheme. This study appears to be definitely beyond the purposes of
this paper.
Still, simply to perform a pioneering investigation, we have examined a first
example of such a proposal. to be definite, we have chosen the recently
proposed approach  that can be called {\em light stop scenario} connected with 
electroweak baryogenesis~\cite{Lari}. These scenarios involve CP violating phases. Here, 
we simply exploit some features of the expected mass spectrum. In particular, 
in these models, one of the stop quarks is particularly light (around one hundred GeV)
and one very light neutralino and one very light chargino also exist. In
principle, this might lead to a sensible effect of Yukawa kind, 
coming from the vertex with virtual stop,
chargino and neutralino, which might be satisfactorily parametrized via
a logarithmic
expansion in view of the common smallness of the involved masses. 

A scenario of this kind was already investigated by Hollik's group
in the process of $t\overline t$ production~\cite{HollikTop}, and led to reasonably large
(of the ten percent size) SUSY virtual effects. We took then mass values
of this point and computed the related effect on the distribution,
allowing the light stop mass to vary between 105 and 120 GeV and fixing the
remaining parameters as in~\cite{HollikTop}.
Fig.~(\ref{fig:Hollik}) shows the complete relative effect at 1 TeV. As one sees, 
at 1 TeV and away from the threshold peak, the situation is quite similar to that already
discussed with the genuine SUSY effect giving a (rather) small positive contribution of a few percents.
Instead, near the threshold there can be a strong peak that could be visible at the expected LHC experimental accuracy.
It is reasonable to guess that after dedicated analysis of these threshold effects, possibly including width
or higher order effects, some large effect could survive in the neighborhood of the threshold.

As a technical remark, we warn the reader that in this scenario, the light
neutralino is of the Higgsino type. Other points, as for instance 
the Les Houches 2005 benchmark point defined as LHS-2, recently proposed to experimental consideration at 
LHC~\cite{LariTalk}, have a bino-like light neutralino which depresses the 
above peak effect leaving a $\simeq +2\%$ genuine SUSY effect at 1 TeV mildly dependent on the 
light stop mass.

Given the rather vague theoretical motivations of our choice, we consider this
result as a purely indicative one. Still, it seems to indicate that large
SUSY effects in the process might arise from symmetry
breaking schemes less constrained than the simplest mSUGRA choice. 
A more rigorous analysis of this possibility will be performed in a following paper.

\section{Conclusions}

We have performed in this paper the complete calculation of the one-loop
electroweak contribution to the process of $t$-channel single top production
in the MSSM, with mSUGRA symmetry breaking scheme. We have found that, for
general choices of benchmark points, the genuine SUSY effect appears to
be, at the expected level of experimental accuracy, hardly visible, {\em i.e.} at the
few percent level. An exception to this statement might be represented by
threshold effects occurring at particular points of the 
MSSM parameter space. These points can easily be allowed in 
symmetry breaking schemes more general than mSUGRA.
As an illustrative example, we have discussed a MSSM configuration involving 
a light stop squark and a light neutralino, in vicinity of the threshold $m_t = m_{\widetilde t}+
m_{\chi^0}$. 
This possibility requires, though, a deeper investigation beyond the purposes of this paper.

The main conclusion of the paper comes in fact from the calculation of the
conventional Standard Model one-loop electroweak effect. We have shown that
its value is (unexpectedly) large, reaching the 10\% size in the
total rate, computed over a realistic invariant mass range. 
This value is well competitive with that of the available NLO QCD calculations, and must
therefore be accurately retained and taken into proper account in any
dedicated future program that aims to provide predictions for rates beyond
the simplest perturbative Born level~\cite{Frixione:2005vw}.

The fact that the NLO electroweak effect is competitive with the NLO QCD one
appears to us, indeed, a unique feature of this process of single top production. For this 
reason we would like to expand this point. {\em A priori}, it is nowadays well known that in the high-energy regime, the asymptotic
electroweak  corrections are dominated at one-loop by large squared logarithms $\alpha\, \log^2(s/M_W^2)$
where $\sqrt{s}$ is the typical energy scale of the process. At LHC, these large logarithms can 
enhance the electroweak correction and easily reach the size of NLO QCD corrections.
However, this kind of analysis is rather qualitative. In the end,  a complete
one loop calculation is always  required  to determine safely  the actual size of radiative corrections
in realistic energy ranges, not necessarily asymptotic.

Examples of LHC processes where the full one-loop calculation reveals indeed large 
electroweak corrections are for instance weak corrections onto b-jet, prompt-photon, and $Z$-production~\cite{Maina:2004ve}.
On the other hand, for single top production processes, our previous analysis of the 
associated $tW$ production~\cite{Beccaria:2006dt} showed that the electroweak corrections to the 
integrate cross section are typically well below 10\%.
Instead and remarkably, the $t$-channel process has corrections which are beyond this value in realistic observables
like the integrated cross section from threshold up to the moderate invariant mass $M_{td}\simeq 500$ GeV.


A final comment should be added concerning the electroweak effects of
supersymmetric physics beyond the Standard Model. In our MSSM analysis we
have found, in general small genuine SUSY effects. However, the overall
size of the one-loop contribution, although essentially produced by the SM
component, remains large and observable and, in some rather special cases,
it could exhibit a peak.  In this sense it seems to us that, for the cases
that we have considered, one can indeed consider the $t$-channel single top
production process as a realistic test of electroweak models and possibly,
as in the original definition of~\cite{Tait:1999cf,Yuan:2006ck} {\em a window to new physics}.

\newpage
\appendix

\section{Sudakov expansion of the process $ub\to td$ in the SM and MSSM}

\subsection{Sudakov expansion from general rules}

The Born amplitude can be written with explicit helicity quantum numbers of the
external fermions
\bq
A^{Born}={2\pi\alpha\over s^2_W(t-M^2_W)}
[\bar u(d,\tau')\gamma^{\mu}P_Lu(u,\lambda')]
[\bar u(t,\tau)\gamma_{\mu}P_Lu(b,\lambda)]
\eq
\noindent  
where $\lambda,\lambda',\tau,\tau'$ are the $b,u,t,d$
helicities,
$P_{R,L}=(1\pm\gamma^5)/2$ are the projectors on
$R,L$ chiralities.

It  is convenient to work with 
helicity amplitudes $F_{\lambda,\lambda',\tau,\tau'}$; 
retaining only the top 
mass and setting
all the remaining masses equal to zero leaves one single 
amplitude $F_{----}$:

\bq
F^{Born}_{----}={4\pi\alpha s \sqrt{\beta}\over s^2_W(t-M^2_W)}
\eq
\noindent
with $\beta={p_t\over E_t}=1-{m^2_t\over s}$.

The expression of the differential cross section after color average is 
\bq
{d\sigma^{Born}\over d\cos\theta}=
{\beta^2\pi\alpha^2 s\over8s^4_W(t-M^2_W)^2}
\eq

At one-loop, the Sudakov electroweak corrections can 
be of universal and of angular dependent kind.
 
The effect of the universal terms on the helicity amplitude
can be summarized as follows:

\bq
F^{Univ}_{----}=
F^{Born}_{----}
~{1\over2}~[~c^{ew}(b\bar b)_L+c^{ew}(u\bar u)_L+
c^{ew}(d\bar d)_L+c^{ew}(t\bar t)_L]
\eq

where, in the MSSM~\cite{Beccaria:2003yn}:

\bq
c^{ew}(q\bar q)_L=c^{ew}(\tilde{q}\tilde{\bar q})_L=
c(q\bar q, ~\mbox{gauge})_L~+~c(q\bar q,~\mbox{Yukawa})_L
\eq

\bq
c(d\bar d, ~\mbox{gauge})_L=c(u\bar u, ~\mbox{gauge})_L={\alpha(1+26c^2_W)\over144\pi
s^2_Wc^2_W}~(2~ \log{s\over M^2_W}-\log^2{s\over M^2_W})
\eq
\bqa
c(b\bar b, ~\mbox{Yukawa})_L&=&
c(t\bar t, ~\mbox{Yukawa})_L= \\
&=& -~{\alpha\over8\pi s^2_W}~
[\log{s\over M^2_W}]~
[{m^2_t\over M^2_W}(1+\cot^2\beta)+{m^2_b\over M^2_W}(1+\tan^2\beta)], \nonumber
\eqa
where $\tan\beta$ is, as usual,  the ratio $v_2/v_1$ of Higgs vacuum expectation values.

The scale of the squared logarithms is determined at this NLO logarithmic order in the Sudakov
expansion. It is always a gauge boson mass. It can be $M_W, M_Z$ or the fictitious IR regulating
photon mass $M_\gamma$. The high-energy $SU(2)\times U(1)$ gauge structure is clearer if we set
write all expressions with $M_\gamma, M_Z$ set to $M_W$. The above expressions adhere to this 
convention. 
For what concerns the single logarithms, 
the scale is arbitrary at logarithmic NLO. Using $M_W$ as the logarithmic 
scale of the expansion, as we do in this
discussion, leaves out  residual NNLO energy independent terms $\sim\log(M/M_W)$, 
where $M$ is the, possibly  different, true scale. We discuss 
this important point in section III.

The angular dependent terms have the following expression:

\bq
F^{ang}_{----} = F^{Born}_{----}\,c^{ang}_{----}
\eq
where
\bq
c^{ang}_{----}=-~{\alpha(1+8c^2_W)\over18\pi s^2_Wc^2_W} 
[\log{-u\over s}][\log{s\over M^2_W}]
-~{\alpha(1-10c^2_W)\over36\pi s^2_Wc^2_W} 
[\log{-t\over s}][\log{s\over M^2_W}]
\eq
At high energy we have $t\simeq-~{s\over2}(1-\cos\theta)$
and $u\simeq-~{s\over2}(1+\cos\theta)$.

In addition to the previous terms of Sudakov type, 
there are at one-loop "known" linear 
logarithms of RG origin, whose expression we quote for completeness:
\bq
F^{RG}_{----}=-~{1\over4\pi^2}[g^4\tilde{\beta^0}
{dF^{Born}_{----}\over dg^2}][\log\frac{s}{M_W^2}] ={\alpha^2 s \sqrt{\beta}\over s^4_W(t-M^2_W)}[\log\frac{s}{M_W^2}]
\eq
using the lowest order Renormalization
Group $\beta$ function for the gauge
coupling $g=e/s_W$: $\tilde{\beta^0}=-~{1\over4}$ in MSSM, $\tilde{\beta_0}=\frac{19}{24}$ in SM.

\subsection{Sudakov expansion from the diagrammatic expansion}

We now list all the separate energy-growing MSSM contributions to the radiatively corrected process.
At the end, we shall combine them to reproduce the previous NLO expansion.

\subsubsection{Born amplitude and corrections}

As we have seen, the asymptotic form of the Born amplitude in the $(-,-,-,-)$ helicity channel is
\be
F^{\rm Born}_{----} = 
\frac{4\pi\alpha}{s_W^2}\frac{s}{t}. 
\ee
In the Sudakov approximation, we obtain the full amplitude by adding several energy growing terms which we shall denote
as 
\be
F^{\rm Sudakov}_{----} = F^{\rm Born}_{----}  + F^{\rm WW}_{----} + F_{----}^{\Delta, \rm light\, quark} + F_{----}^{\Delta, \rm heavy\,quark} + F_{----}^{\Box, \rm direct} 
+ F_{----}^{\Box, \rm twisted}.
\ee
The origin of the various terms is as follows. $F^{\rm WW}_{----}$ comes from the $W$ gauge boson self energy, $F_{----}^{\Delta, \rm light\,quark} 
+ F_{----}^{\Delta, \rm heavy\,quark}$
is the contribution from the vertex corrections, $F_{----}^{\Box, \rm direct} 
+ F_{----}^{\Box, \rm twisted}$ is from the two types of box diagrams.

We now list the various detailed expressions for the corrections.

\subsubsection{W self energy}

The W self-energy contribution to the helicity amplitude is 
\be
F^{\rm WW}_{----} = \frac{5\alpha^2}{s_W^4}\frac{s}{t}\log s
\ee

\subsubsection{Vertex corrections}

The light quark vertex correction is ($M_{Z, W, \gamma}\to M_V$)
\be
F_{----}^{\Delta, \rm light\,quark} = \frac{2\alpha^2}{s_W^4}\frac{s}{t}\left[(2\log s-\log^2\frac{t}{M_V^2})\frac{1-10c_W^2}{72 c_W^2} \right]
\ee
The heavy quark vertex correction is
\be
F_{----}^{\Delta, \rm heavy\,quark} = \frac{2\alpha^2}{s_W^4}\frac{s}{t}\left[(2\log s-\log^2\frac{t}{M_V^2})\frac{1-10c_W^2}{72 c_W^2}
-\frac{1}{4M_W^2}(\widehat{m}_t^2+\widehat{m}_b^2)\log s\right],
\ee
where $\widehat{m}_t = m_t/\sin\beta$ and $\widehat{m}_b = m_b/\cos\beta$.

\subsubsection{Box diagrams}

The box logarithmic terms only arise in the SM. The direct box 
contribution  is
\be
F_{----}^{\Box, \rm direct} = \alpha^2\frac{1-10c_W^2}{9s_W^4 c_W^2}\frac{s}{t}\log^2\frac{s}{M_V^2}
\ee
The twisted box contribution is 
\be
F_{----}^{\Box, \rm twisted} = -\alpha^2\frac{1+8 c_W^2}{9s_W^4 c_W^2}\frac{s}{t}\log^2\frac{u}{M_V^2}
\ee

\subsubsection{Summing up: The complete Sudakov expansion}

We can separate the angular single logarithms as follows
\ba
F_{----}^{\Delta, \rm light+heavy\,quark} &=& F_{----}^{\rm Born}\ \frac{\alpha}{\pi}\left[
\frac{1-10c_W^2}{72s_W^2 c_W^2}(2\log s-\log^2\frac{s}{M_V^2})-\frac{1}{8 M_W^2 s_W^2}(\widehat{m}_t^2+\widehat{m}_b^2)\log s\right. \nonumber \\
&& \left. 
-\frac{1-10c_W^2}{36  s_W^2 c_W^2}\log s\log\frac{t}{s}
\right]
\ea
Also,
\be
F_{----}^{\Box, \rm direct+twisted} = F_{----}^{\rm Born}\ \frac{\alpha}{\pi}
\left[-\frac{1}{2s_W^2}\log^2\frac{s}{M_V^2}-\frac{1+8c_W^2}{18 s_W^2 c_W^2}\log s\log\frac{u}{s}\right]
\ee
Adding and subtracting $-1/s_W^2 \log s$ we can write the factor in square brackets as
\be
\left[\cdots\right] = -\frac{1}{s_W^2}\log s +\frac{1}{2s_W^2}(2\log s -\log^2\frac{s}{M_V^2})-\frac{1+8c_W^2}{18 s_W^2 c_W^2}\log s\log\frac{u}{s}
\ee
Therefore, in conclusion, in the MSSM:
\be
F_{----}^{\rm Sudakov} = F_{----}^{\rm Born} \cdot c + F_{----}^{RG}
\ee
where
\ba
c &=& \frac{\alpha}{\pi}\left[\frac{1+26 c_W^2}{72 s_W^2 c_W^2}(2\log s-\log^2\frac{s}{M_V^2})-\frac{1}{8 M_W^2 s_W^2}(\widehat{m}_t^2+\widehat{m}_b^2)\log s \right. \nonumber \\
&& \left. -\frac{1+8c_W^2}{18 s_W^2 c_W^2}\log s\log\frac{u}{s}-\frac{1-10c_W^2}{36  s_W^2 c_W^2}\log s\log\frac{t}{s}\right] \\
F_{----}^{RG} &=& \frac{\alpha^2}{s_W^4}\frac{s}{t}\log s
\ea
The RG log is a combination of the added/subtracted single logarithm plus the WW term
\be
F_{----}^{RG} = F_{----}^{\rm Born}\ \frac{\alpha}{\pi}\left(-\frac{1}{s_W^2}\log s\right) + F_{----}^{\rm WW}
\ee

In the SM there are changes in the triangles and in the WW self energy. The final result is quite similar
and reads
\ba
c &=& \frac{\alpha}{\pi}\left[\frac{1+26 c_W^2}{72 s_W^2 c_W^2}(3\log s-\log^2\frac{s}{M_V^2})-\frac{1}{16 M_W^2 s_W^2}(m_t^2+m_b^2)\log s \right. \nonumber \\
&& \left. -\frac{1+8c_W^2}{18 s_W^2 c_W^2}\log s\log\frac{u}{s}-\frac{1-10c_W^2}{36  s_W^2 c_W^2}\log s\log\frac{t}{s}\right] \\
F_{----}^{RG} &=& -\frac{19}{6}\frac{\alpha^2}{s_W^4}\frac{s}{t}\log s
\ea

as one sees, these results are in full agreement with the expansion obtained from general rules in~\cite{Beccaria:2003yn}.

\begin{table}
\begin{tabular}{|l|llll|}
\hline
                        &SU1                &SU6        &LS1 &LS2 \\
\hline
$m_0$                    & 70            & 320         & 300              & 300\\
$m_{1/2}$                & 350           & 375         & 150              & 150\\
$A_0$                    & 0             & 0           & -500             & -500\\
$\tan\beta$              & 10            & 50          & 10               &  50\\
$\mu/|\mu|$              & 1             & 1           & 1                &  1\\
$\alpha$                 & -0.110       & -0.0212	 & -0.109	     &      -0.015\\
$M_1$                    & 144.2  	 &     155.8  &  	 60.1	      &    60.6\\
$M_2$                    & 270.1  	  &    291.3   & 	114.8	       &   115.9\\
$\mu$                    & 474.4  	   &   496.6  	 & 329.7	   & 309.3\\
$H^\pm$                  & 534.3  	    &  401.7  	 & 450.4	   & 228.9\\
$H^0$                    & 528.3  	     & 392.5  	 & 442.5	   & 211.1\\
$h^0$                    & 114.6  	     & 115.7  	 & 111.4	   & 110.8\\
$A^0$                    & 527.9  	     & 392.5  	 & 443.4	   & 212.0\\
$\chi^\pm_1$             & 262.8	     & 289.3	 & 108.0	           & 111.1	 \\
$\chi^\pm_2$	         & 495.3	     & 514.8	 & 350.1	 	   & 329.4	 \\
$\chi^0_1$	         & 140.1	     & 153.0	 & 57.38	 	   & 58.92	 \\
$\chi^0_2$	         & 263.1	     & 289.4	 & 108.5	 	   & 111.3	 \\
$\chi^0_3$	         & 479.2	     & 501.0	 & 335.3	 	   & 315.8	 \\
$\chi^0_4$	         & 495.4       &  514.0    	 & 348.7    	   & 326.5    \\
\hline
\end{tabular}
\quad\quad\quad
\begin{tabular}{|l|llll|}
\hline
                        &SU1                &SU6        &LS1 &LS2 \\
\hline
$\widetilde{l}_L$        & 253.3  	     & 412.3  	 & 321.0	   & 321.2\\
$\widetilde{l}_R$        & 157.6  	     & 353.4  	 & 308.7	   & 308.7\\
$\widetilde{\nu}_e$      & 241.0  	     & 404.8  	 & 311.3	   & 311.3\\
$\widetilde{\tau}_L$     & 149.6  	     & 195.8  	 & 297.1	   & 078.1\\
$\widetilde{\tau}_R$     & 256.1        & 399.2  &  	323.8	   & 282.5\\
$\widetilde{\nu}_\tau$   & 240.3        & 362.5  &  	308.4	   & 243.6\\
$\widetilde{u}_L$        & 762.9        & 870.5   & 	459.8	   & 460.2\\
$\widetilde{u}_R$        & 732.9  	 &     840.7  &  	451.9	 &   452.3\\
$\widetilde{d}_L$        & 766.9  	 &     874.0   & 	466.4	  &  467.0\\
$\widetilde{d}_R$        & 730.2  	 &     837.8  	 & 452.8	   & 453.2\\
$\widetilde{t}_L$        & 562.5  	 &     631.5  	 & 213.3	   & 223.6\\
$\widetilde{t}_R$        & 755.8  	 &     796.9  	 & 462.9	   & 431.3\\
$\widetilde{b}_L$        & 701.0  	 &     713.7  	 & 380.6	   & 304.0\\
$\widetilde{b}_R$        & 730.2  	 &     787.6  	 & 449.1	   & 401.7\\
$\theta_\tau$            & 1.366        & 1.133   	 & 1.091	           & 1.117  \\
$\theta_b$               & 0.3619       & 0.7837  	 & 0.184	           & 0.653 \\
$\theta_t$               & 1.070        & 1.050   	 & 1.016             &  0.9313 \\
\hline
\end{tabular}
\caption{
Table of spectra for the various benchmark points. All entries with the dimension of a mass are expressed in GeV. 
The spectra have been computed with the code SUSPECT~\cite{Djouadi:2002ze}.}
\label{tab:spectra}
\end{table}

\newpage


\newpage

\newpage

\newpage

\newpage

\begin{figure}[tb]
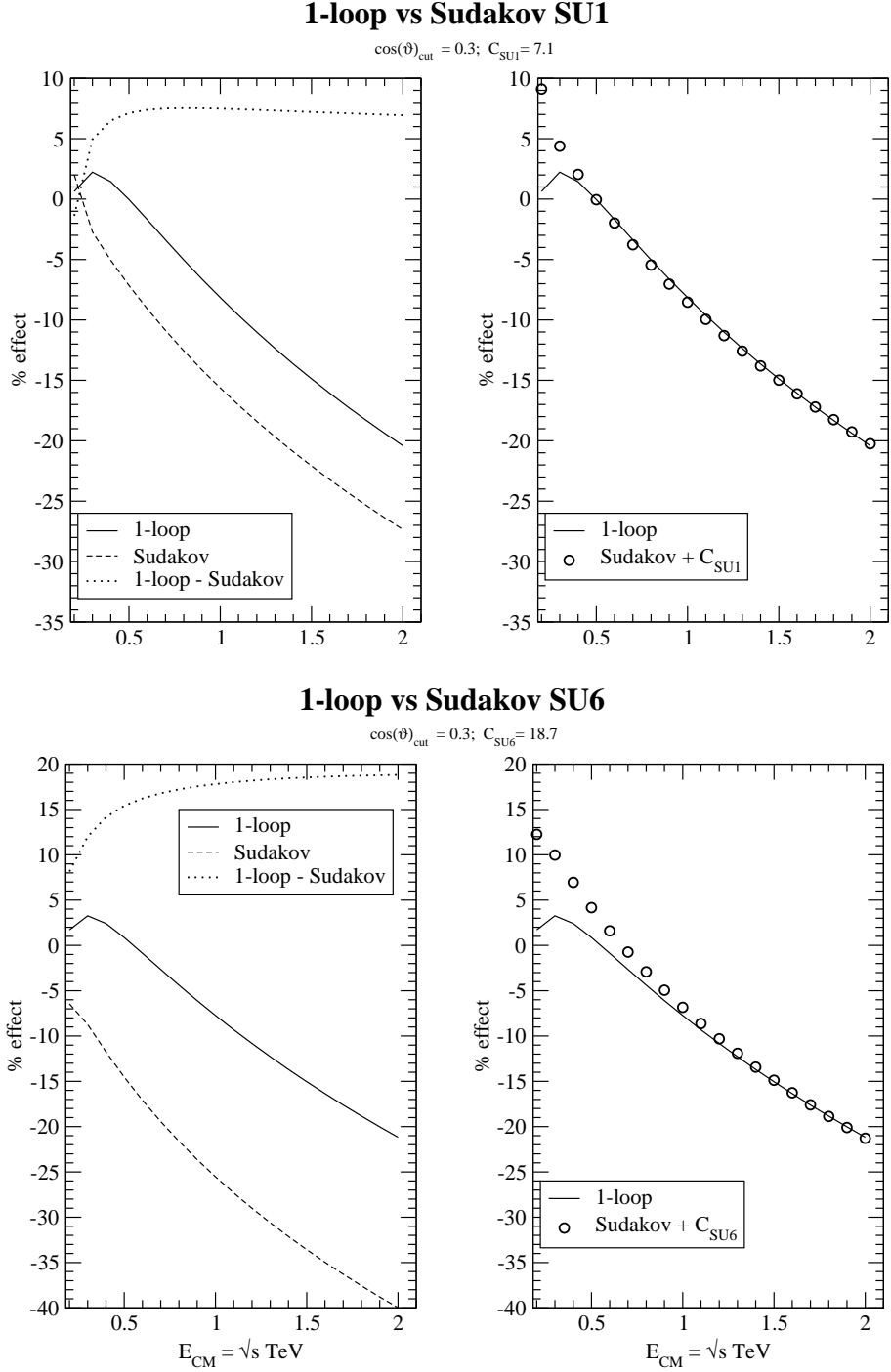

\centering
\epsfig{file=Figures/SudakovSU1.eps, width=12cm, angle=0}
%
\epsfig{file=Figures/SudakovSU6.eps, width=12cm, angle=0}
\vspace{1.5cm}
\caption{Comparison between the full one-loop calculation in the MSSM and the NLO Sudakov approximation.
Two scenarios are considered, SU1 and SU6. As in the SM case, 
a strong angular cut is imposed as well as the fictitious definition $M_\gamma = M_Z$. Real QED radiation
is consistently switched off.}
\label{fig:sudakovMSSM}
\end{figure}

\newpage

\begin{figure}[tb]
\centering
\epsfig{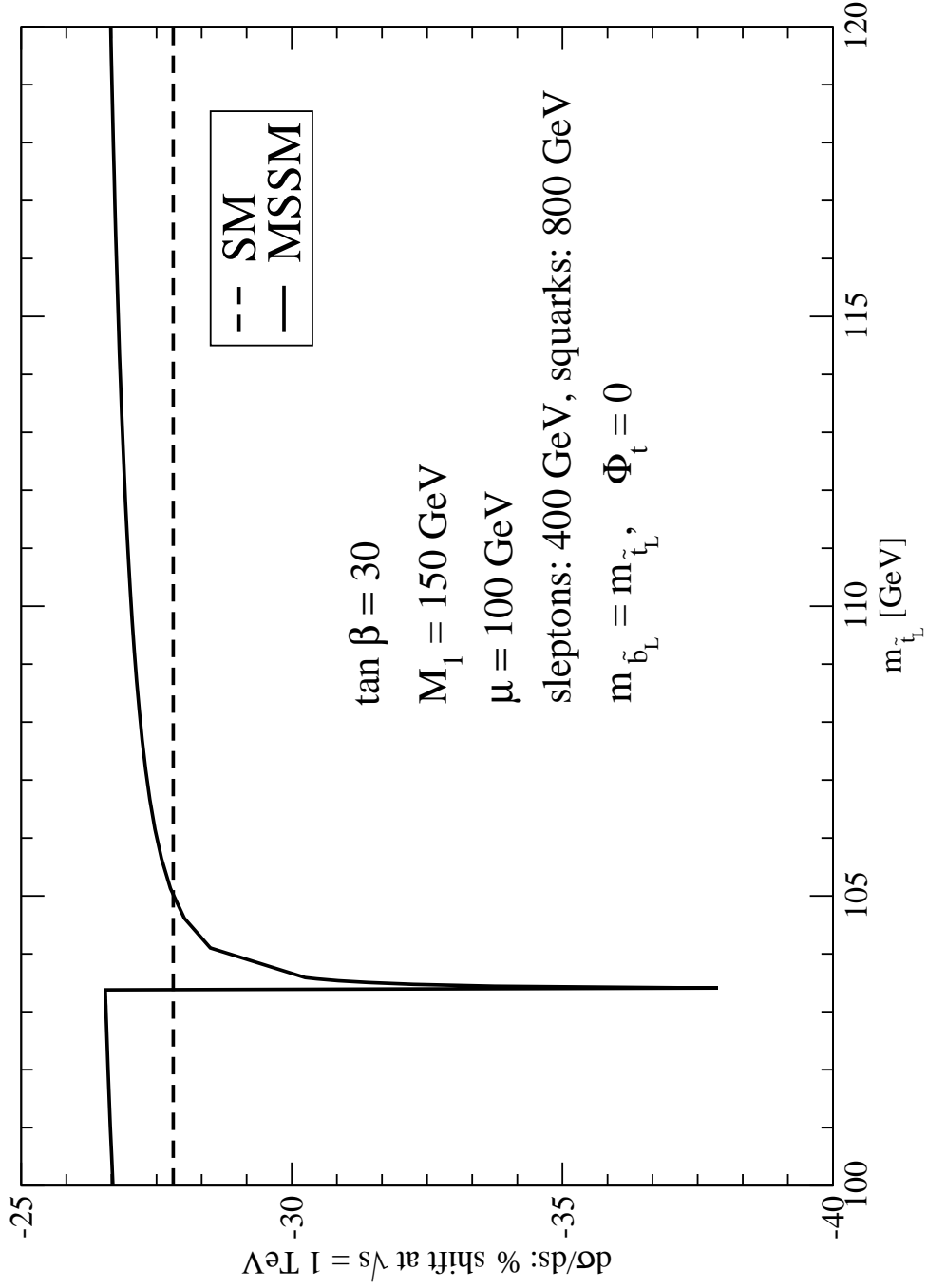}
\vspace{1.5cm}
\caption{Threshold effect in the light stop scenario. The position of the peak is the point where the sum of 
light stop and light neutralino masses sum up to the top quark mass. The dashed line is the (constant) SM value.
The effects are computed at the fixed reference value $\sqrt{s}=$ 1 TeV.}
\label{fig:Hollik}
\end{figure}


\end{document}